\documentclass{article}
\newcommand{\bfr}{\begin{flushright}}
\newcommand{\efr}{\end{flushright}}
 
\begin{document}
\title{Quantum Corrections to Entropy of Charged Dilatonic Black Holes
in Arbitrary Dimensions
}
\author{Kiyoshi Shiraishi\\
Akita Junior College, Shimokitade-Sakura, Akita-shi, \\Akita 010,
Japan
}
\date{Modern Physics Letters {\bf A9} (1994) pp. 3509-3516
}
\maketitle
\begin{abstract}
The quantum contribution of a scalar field to entropy of a dilatonic
black hole is calculated in the brick wall model by the WKB method and
analyzed by a high-temperature expansion. If the cutoff distance from
the horizon approaches zero, the leading divergent piece of entropy
turns out to be proportional to the ``area'' of the horizon surface
(which has $(N-1)$-dimensional extension in $(N+1)$-dimensional
space-time) and independent of other properties of black holes even in
the case of general dilaton coupling. There is also qualitative
argument with the known result of subleading divergence for $N=3$.\\
PACS Nos.: 04.60.+n, 12.10.Gq, 97.60.Lf, 11.10.Kk.
\end{abstract}

\bigskip

Recently, there has been a renewed interst in thermodynamic description
of black holes. The ``zero-loop'' partition function of black holes is
derived from the path-integral of a classical action of Einstein
gravitation, along with a special attention to the boundary
conditions.\cite{1}

An attempt to calculate contribution of quantum fields to the black hole entropy
was made in the seminal work of 't Hooft.\cite{2} In the last decade,
entropy due to quantum fields and string excitations has been
investigated by many authors. Recently, Callan and Wilzcek analyzed
field theory in an accelerated system and obtained quantum contribution
to the entropy, which they called as geometric entropy.\cite{3} Their
geometric entropy is believed to be equivalent to the one obtained in
the other ways such as was done in Ref. \cite{4}. Susskind and
Uglum \cite{5} and Kabat and Strasslers \cite{6} showed that similar
result can be obtained through the WKB method, which is used in the 't
Hooft's analysis.\cite{2} 

In a recent paper,\cite{7} Ghosh and Mitra calculated the
entropy due to quantum scalar field in the background of a dilatonic
black hole in effective string theory by the WKB approximation. They
showed the cutoff-dependent part is not proportional to the area of
the horizon surface if the black hole is maximally charged. Recent
calculation by Solodukhin \cite{8} supports their evaluation. 

In this letter,
we compute quantum corrections to black hole entropy in a model with an
arbitrary dilaton coupling in arbitrary space-time dimensions using the
WKB method. First of all, we start with reviewing the classical
background of a charged dilatonic black hole. Suppose the classical
action for gravity with a U(1) gauge field $A_\mu$ and a dilaton field
as follows: 
\begin{equation}
S =\int d^{N+1}x \frac{\sqrt{-g}}{16\pi}\left[R-
\frac{4}{N-1}(\nabla\phi)^2 - e^{- 4a\phi/( N -1)}
F^2\right]+\mbox{(boundary terms)}\,,
\label{eq1}
\end{equation}
where $F^2 = F^{\mu\nu}F_{\mu\nu}$ with the field strength
$F_{\mu\nu}=\partial_\mu A_\nu-\partial_\nu A_\mu$ and the Newton
constant is set to unity. Here we assume that the dilaton coupling a
may take any positive value: For effective field theories of string
theory, it takes $a = 1$.

The spherical black hole solution to the field equation derived from the
action (\ref{eq1}) is given by \cite{9,10}
\begin{equation}
ds^2=-\sigma^{-2}(r)\Delta(r)dt^2+\sigma^{2/(N-2)}(r)
\left[\frac{dr^2}{\Delta(r)}
+r^2d\Omega^2_{N-1}\right]\,,
\label{eq2}
\end{equation}
with
\begin{equation}
\Delta(r)=\left[1-\left(\frac{r_+}{r}\right)^{N-2}\right]
\left[1-\left(\frac{r_-}{r}\right)^{N-2}\right]\,,
\end{equation}
\begin{equation}
\Delta(r)=
\left[1-\left(\frac{r_-}{r}\right)^{N-2}\right]^{a^2/(N
-2+a^2)}\,,
\end{equation}
where $r_+$ and $r_-$ are constants $(r_+>r_-)$. In
Eq.~(\ref{eq2}), $d\Omega^2_{N-1}$ represents the line element of a
unit $(N-1)$-sphere.

For this solution, the dilaton field configuration is expressed as
\begin{equation}
e^{4a\phi/(N-1)}=\sigma^2(r)
\end{equation}
and the electric field is written as
\begin{equation}
F=\frac{Q}{r^{N-1}}dt\wedge dr\,,
\end{equation}
where the charge $Q$ is related to $r_+$ and $r_-$ as
\begin{equation}
Q^2=\frac{(N-1)(N-2)^2}{2(N-2+a^2)}(r_+r_-)^{N-2}\,.
\end{equation}
The mass $M$ of the black hole described by the solution is given by
\begin{equation}
M=\frac{(N-1)A_{N-1}}{16\pi}\left(r_+^{N-2}+
\frac{N-2-a^2}{N-2+a^2}r_-^{N-2}\right)\,,
\end{equation}
where
\begin{equation}
A_{N-1}=\frac{2\pi^{N/2}}{\Gamma(N/2)}\,.
\end{equation}

Now we introduce an extra neutral scalar field $\varphi$ with a mass $m$
in the black bole background. The scalar is allowed to be coupled with
the background dilaton field in general. We will evaluate the
contribution of this field to black hole entropy by means of a mode
function technique. The equation of motion of the scalar field can be
written in the general form:
\begin{equation}
\nabla^\mu e^{-4ab\phi/(N-1)}\nabla_\mu\varphi=
e^{-4ac\phi/(N-1)}m^2\varphi\,,
\label{eq10}
\end{equation}
where $b$ and $c$ are parameters representing the strength of the
coupling to the dilaton field.

In the spherically symmetric space, the scalar field can be written by
a product of mode functions:
\begin{equation}
\varphi=\frac{1}{r^{(N-1)/2}}e^{-i\omega t}\chi_{\omega\ell}(r)
Y^{(N-1)}_{\ell\nu}(\Omega)\,,
\label{eq11}
\end{equation}
where the generalized spherical function $Y^{(N-1)}_{\ell\nu}(\Omega)$
is the eigenfunction of the Laplacian on $S^{N-1}$ with a unit radius,
such that 
\begin{equation}
\Delta^{(N -1)}Y^{(N
-1)}_{\ell\nu}=-\ell(\ell+N-2)Y_{\ell\nu}^{(N-1)}\,,
\quad (\ell=0,1,2,\dots)\,.
\end{equation}

The differential equation concerning the radial function
$\chi_{\omega\ell}(r)$ can be obtained by substituting Eq.~(\ref{eq11})
into Eq.~(\ref{eq10}). We further introduce a new radial coordinate $y$
defined by \cite{10}
\begin{equation}
y=\int^r \frac{\sigma^{2b}}{\Delta}dr\, , 
\end{equation}
where $y$ varies between $-\infty$ and $\infty$ while $r$ varies
between
$r_+$ and $\infty$. Then the wave equation for the radial function is
simplified as \cite{10}
\begin{equation}
\frac{d^2\chi_{\omega\ell}}{dy^2}+W\chi_{\omega\ell}=0\, ,
\label{eq14}
\end{equation}
where
\begin{eqnarray}
W(y)&=&\left(\frac{\Delta(r)}{\sigma^{2b}(r)}\right)^2
\frac{\sigma^{2/(N-2)}(r)}{\Delta(r)}\left[
\frac{\sigma^2(r)}{\Delta(r)}\omega^2-
\frac{\ell(\ell+N-2)}{r^2\sigma^{2/(N-2)}(r)}\right.\nonumber
\\ &
&\left.-\frac{N-1}{2r\sigma^{2/(N-2)}(r)}\frac{d\Delta(r)}{dr}
-m^2\sigma^{2(b-c)}(r)-\Delta(r)\mu^2(r)\right]\,,
\end{eqnarray}
with
\begin{equation}
\mu^2(r)=\frac{1}{\sigma^{2/(N-2)}(r)}\left[\frac{(N-1)(N-3)}{4r^2}
-\frac{N-1}{r}b\frac{1}{\sigma(r)}\frac{d\sigma(r)}{dr}\right]\,.
\end{equation}

Let us use the WKB method to evaluate the solution to Eq.~(\ref{eq14}).
The solution is approximated as
\begin{equation}
\chi_{\omega\ell}(y)\approx\frac{1}{\sqrt{W(y)}}\int^y  
\sin\left(\sqrt{W(y)}\right) dy\,.
\end{equation}
If we impose the boundary condition that the wave function vanishes on
the ``brick wall'' at $r = r_+ + \delta$ and on the wall of a large box
at $r=L$,\cite{2} we find that the discrete quantum number $n$ is
associated with the phase function \cite{2,5,6,7,11}:
\begin{equation}
n\pi=\int_{r_++\delta}^L  k_{\omega\ell}(r)dr\, ,
\end{equation}
where
\begin{eqnarray}
&&k_{\omega\ell}(r)\nonumber \\
&&\!\!\!\!\!\!\!\!\!\!\!\!\!\!\!\!=\sqrt{\frac{\sigma^{2/(N-2)}(r)}{\Delta(r)}\left[
\frac{\sigma^2}{\Delta}\omega^2-
\frac{\ell(\ell+N-2)}{r^2\sigma^{2/(N-2)}}-\frac{N-1}{2r\sigma^{2/(N-2)}}
\frac{d\Delta}{dr}-m^2\sigma^{2(b-c)}-\Delta\mu^2\right]}\,.
\end{eqnarray}

The free  energy $F$  can  be  derived  from  the  density  of  the
modes.   After  some manipulation,\cite{2,5,6,7,11}  one  can  get  the 
following  expression  for  the  free  energy  at inverse temperature
$\beta$:
\begin{equation}
F=-{\frac{1}{\pi }} \sum\limits_{\ell =0}
^{\infty } {D}_{\ell }\int_{}^{}d\omega 
{\frac{1}{{e}^{\beta \omega }-1}}\int_{{r}_{+}+\delta }
^{L}{k}_{\omega\ell }(r)dr \equiv  \int_{{r}_{+}+\delta }^{L}f(r)dr\,,
\label{eq20}
\end{equation}
where the degeneracy of the spherical modes $D_{\ell}$ is \cite{12}
\begin{equation}
{D}_{\ell }\equiv {\frac{(2\ell +N-2)(\ell +N-3)!}{(N-2)!\ell !}}\,.
\end{equation}
In (\ref{eq20}), the integral over $\omega$ should be taken only over
possible  values for 
which $k_{\omega\ell}$ is real.

We wish to study the portion that yields the divergence of the free energy 
as $\delta \rightarrow 0$. In the first step to this end,
 we approximate the ``density'' $f(r)$. 
Integrating over $\omega$, $f(r)$ can be rewritten by using the modified Bessel 
function \cite{13} as
\begin{equation}
f(r)=-{\frac{{\sigma }^{-(N-3)/(N-2)}}
{\pi }} \sum\limits_{\ell =0}^{\infty } 
{D}_{\ell }\sum\limits_{p=1}^{\infty } 
{\frac{\eta (r)}{p\beta \sqrt {\Delta {\sigma }^{-2}}}}{K}
_{1} (p\beta \sqrt {\Delta {\sigma }^{-2}}\eta (r))\,,
\label{f}
\end{equation}
where $\eta (r)$ is defined by
\begin{equation}
\eta (r)=\sqrt {{\frac{\ell (\ell +N-2)}{{r}^{2}{\sigma }^{2/(N-2)}}}+{\frac{N-1}
{2r{\sigma }^{2/(N-2)}}}{\frac{d\,\Delta }{d\,r}}+{m}^{2}
{\sigma }^{2(b-c)}+\Delta {\mu }^{2}}\,.
\end{equation}
Note that the inverse temperature $\beta$ appears only in the
combination 
$\beta \sqrt {\Delta {\sigma }^{-2}}
=\beta \sqrt {\left|{{g}_{tt}}\right|}\equiv 
\tilde{\beta }(r)$. To see the structure of generating divergences in 
free energy, we will expand Eq.~(\ref{f}) with respect to powers of 
$\tilde{\beta }(r)$; this is just a ``high-temperature'' expansion. 
To this end, we 
use another integral representation of the modified Bessel
function.\cite{13} Then  Eq.~(\ref{f}) becomes
\begin{equation}
f(r)=-{\frac{{\sigma }^{-(N-3)/(N-2)}}{\pi }}
\sum\limits_{\ell =0}^{\infty } 
{D}_{\ell }\sum\limits_{p=1}^{\infty }
 {\frac{1}{{p}^{2}{\tilde{\beta }}^{2}}}\int_{0}^{\infty }
\exp\left({-t-{\frac{{p}^{2}{\tilde{\beta }}^{2}{\eta }^{2}}
{4t}}}\right)dt\,.
\label{ff}
\end{equation}
To perform the sum over $\ell$, we use the following expansion
\cite{12} in Eq.~(\ref{ff}):
\begin{equation}
\sum\limits_{\ell =0}^{\infty } {D}_{\ell }{ e}^{-\ell (\ell +N-2)x}
={\frac{\Gamma  ((N-1)/2)}{\Gamma (N-1)}}{x}^{-(N-1)/2}\left[{1+{\frac{(N-1)(N-2)}{6}}x
+O({x}^{2})}\right]\,.
\end{equation}
Consequently, we find the following expression to the next-to-leading
order for  small $\tilde{\beta }(r)$, under the assumption of
$m{\sigma}^{b-c} \ll {\tilde{\beta }}^{-1}$
 and $r{\sigma }^{1/(N-1)}\gg \tilde{\beta }$:
\begin{eqnarray}
f(r)&=&-\frac{A_{N-1}r^{N-1}\sigma^{2/(N-2)}}{\pi^{(N+1)/2}}
\left\{\frac{1}{\tilde{\beta}^{N+1}}
\Gamma((N+1)/2) \zeta (N+1) \right.
\nonumber \\ 
&+&\frac{1}{4\tilde{\beta}^{N-1}} \Gamma((N-1)/2) \zeta(N-1)
\left[\frac{(N-1)(N-2)}{6r^2\sigma^{2/(N-2)}}-\frac{N-1}
{2r\sigma^{2/(N-2)}}\frac{d\Delta}{dr}\right.\nonumber \\
& &\left.\left.-m^2\sigma^{2(b-c)}-\Delta\mu^2\right]
\right\}+O({\tilde{\beta}}^{-(N-3)})\,,
\end{eqnarray}
where $\zeta(z)$ is the Riemann zeta function and 
we have applied formulas for the gamma functions \cite{13}.

In the second step, we examine the divergence of free energy after 
integration over $r$ as $\delta\rightarrow 0$. We first focus our 
attention on the contribution 
from the leading term of the high-temperature expansion.

The estimation can be done in a straightforward way. To rearrange the 
result, we prepare the following quantities.
The invariant distance between the horizon at $r=r_{+}$ and
$r=r_{+}+\delta$ is \cite{2}
\begin{eqnarray}
& &\varepsilon={2\left({1-{\frac{{r}_{-}^{N-2}}{{r}_{+}
^{N-2}}}}\right)}^{{a}^{2}/[(N-2)(N-2+{a}^{2})]-1
/2}\sqrt {{\frac{{r}_{+}\delta }{ N-2}}}
\nonumber \\ 
& &\times \left\{{1+{\frac{1}{6}} {\frac{ \delta }{{ r}_{ +}}}
\left[{{\frac{  N-1}{2}}  +\left({{\frac{2{a}^{2}}
{N-2+{a}^{2}}}-(N-2)}\right){\frac{{\frac{{r}_{-}^{N-2}}
{{r}_{+}^{N-2}}}}{1-{\frac{{r}_{-}^{N-2}}{{r}_{+}^{N-2}}}}}}
\right]  +O\left({{\left({\delta /{ r}_{ +}}\right)}^{2}}
\right)}\right\}\,.
\end{eqnarray}

The Hawking temperature of the dilatonic black hole is given by 
\cite{9,10}
\begin{equation}
{T}_{H}={\beta }_{H}^{-1}={{\frac{N-2}
{4 \pi {  r}_{  +}}}\left({1-{\frac{{r}_{-}^{N-2}}
{{r}_{+}^{N-2}}}}\right)}^{[{(N-2)}^{2}-{a}^{2}]/[(N-2)(N-2+{a}^{2})]}.
\end{equation}

Then we find that the leading term of the divergent part as 
$\delta \rightarrow 0$, or 
equivalently as $\varepsilon \rightarrow 0$, 
in the free energy of the order of $\beta^{-(N+1)}$ is expressed 
as:
\begin{equation}
{F}_{div1}=-{\frac{1}{(N-1){(2 \pi   )}^{N}}}
{\frac{ \Gamma   ((N+1)/2) \zeta   (N+1)}
{{ \pi }^{(N+1)/2}}} {\frac{{\tilde{  A}}_{  H}
^{  (N-1)}}{{ \varepsilon }^{  N-1}}}
{\frac{{ \beta }_{  H}^{  N}}{{ \beta }
^{  N+1}}}\,,
\end{equation}
where the ``area'' of the horizon $(N-1)$-sphere is given by
\begin{equation}
{\tilde{A}}_{H}^{(N-1)} \equiv   {A}_{N-1}
{\tilde{r}}_{+}^{N-1}\,,
\end{equation}
with
\begin{equation}
{\tilde{r}}_{+} \equiv   {r}_{+}{\sigma }^{1/(N-2)}
({r}_{+}).
\end{equation}
The corresponding entropy turns out to be
\begin{equation}
{S}_{div1}={\beta }^{2}{\frac{ \partial   \,{F}_{div1}}
{ \partial   \,\beta }}={\frac{N+1}{(N-1){(2 \pi
   )}^{N}}}{\frac{ \Gamma   ((N+1)/2) \zeta 
  (N+1)}{{ \pi }^{(N+1)/2}}} 
{\frac{{\tilde{  A}}_{  H}^{  (N-1)}}
{{ \varepsilon }^{  N-1}}}{\left({{\frac{{ \beta }
_{  H}}{{ \beta }^{}}}}\right)}^{  N}.
\label{Sdiv1}
\end{equation}
Notice that the last expression (\ref{Sdiv1}) includes no explicit 
dependence on the 
parameters of the charged dilatonic black hole, $r_{+}$, $r_{-}$, and 
the dilaton 
coupling $a$. Thus the same expression holds for a Schwarzschild or a 
Reissner-Nordstr\"om black hole. Our result is in accord 
with the one in Refs.~\cite{2} and \cite{5} 
and the other reports for $N=3$. Moreover, since the entropy at the
``tree level'' is also proportional to the ``area'' of the horizon
surface, i.e.,
\begin{equation}
{S}_{0}={\frac{1}{4}}{\tilde{A}}_{H}^{(N-1)},
\end{equation}
the contribution (\ref{Sdiv1}) to the entropy can be treated by 
``renormalization'' of the gravitational constant \cite{5,8}.

Next, we turn to the subleading divergences (as $\varepsilon\rightarrow 0$). 
The divergences of the 
order of $\varepsilon^{-(N-3)}$ ($N\ne 3$) is present in both the terms of 
$O({\beta}^{-(N+1)})$ and $O({\beta}^{-(N-1)})$. For $N=3$, these 
divergences are proportional to the logarithm of $\varepsilon$.

For $N\ne 3$, we find
\begin{eqnarray}
& &{F}_{div2}\nonumber \\
&=&-{\frac{(N-2){A}_{N-1}}{12(N-3)
{(2 \pi   )}^{N}}}{\frac{ \Gamma   ((N+1)/2)
 \zeta   (N+1)}{{ \pi }^{(N+1)/2}}}
{\frac{{\tilde{r}}_{+}^{N-3}}{{\varepsilon }^{N-3}}} 
{\frac{{ \beta }_{  H}^{  N}}{{ \beta }^{  N+1}}}
\nonumber \\ 
&\times& \left[{  (N-1)(N+3)\left({1-
{\frac{{r}_{-}^{N-2}}{{r}_{+}^{N-2}}}}\right)+
\left({{\frac{(3{N}^{2}-2N-3){a}^{2}}{N-2+{a}^{2}}}-
2N(N-2)}\right){\frac{{r}_{-}^{N-2}}{{r}_{+}^{N-2}}}}\right]
 \nonumber \\ 
&-&{\frac{{A}_{N-1}}{4(N-3){(2 \pi   )}
^{N-2}}}{\frac{ \Gamma   ((N-1)/2) \zeta 
  (N-1)}{{ \pi }^{(N+1)/2}}}{\frac{{\tilde{r}}_{+}
^{N-3}}{{\varepsilon }^{N-3}}} {\frac{{ \beta }_{  H}
^{  N-2}}{{ \beta }^{  N-1}}} 
\nonumber \\ 
&\times& \left[{{\frac{  (N-1)(N-2)}{6}}  -{\frac{(N-1)
(N-2)}{2}}\left({1-{\frac{{r}_{-}^{N-2}}{{r}_{+}^{N-2}}}}
\right)-{m}^{2}{\tilde{r}}_{+}^{2}{ \sigma }^{2(b-c)}({r}_{+})}
\right]\,,
\end{eqnarray}
and for $N=3$ we obtain
\begin{eqnarray}
{F}_{div2}=&-&{\frac{1}{720}} 
{\frac{{ \beta }_{  H}^{  3}}{{ \beta }^{  4}}}
  \left[{2-{\frac{3}{1+{a}^{2}}}{\frac{{r}_{-}}{{r}_{+}}}}
\right]\ln\left({{\frac{{\Lambda }^{2}}{{\varepsilon }^{2}}}}\right)
\nonumber \\ 
&-&{\frac{1}{72}} {\frac{{ \beta }_{  H}}{{ \beta }
^{  2}}}  \left[{1-3\left({1-{\frac{{r}_{-}}
{{r}_{+}}}}\right)-3{m}^{2}{\tilde{r}}_{+}^{2}{\sigma }^{2(b-c)}
({r}_{+})}\right]\ln\left({{\frac{{\Lambda }^{2}}{{\varepsilon }^{2}}}}
\right),
\end{eqnarray}
where $\Lambda$ is IR cutoff. 

These lead to the following contributions to the entropy:
\begin{eqnarray}
& &{S}_{div2}\nonumber \\
&=&{\frac{(N+1)(N-2){A}_{N-1}}{12(N-3)
{(2 \pi   )}^{N}}}{\frac{ \Gamma   ((N+1)/2)
 \zeta   (N+1)}{{ \pi }^{(N+1)/2}}}
{\frac{{\tilde{r}}_{+}^{N-3}}{{\varepsilon }^{N-3}}} 
{\left({{\frac{{ \beta }_{  H}}{{ \beta }^{}}}}\right)}^{  N}
\nonumber \\ 
&\times&\left[{  (N-1)(N+3)\left({1-{\frac{{r}_{-}
^{N-2}}{{r}_{+}^{N-2}}}}\right)+\left({{\frac{(3{N}^{2}-2N
-3){a}^{2}}{N-2+{a}^{2}}}-2N(N-2)}\right){\frac{{r}_{-}
^{N-2}}{{r}_{+}^{N-2}}}}\right]  
\nonumber \\ 
&+&{\frac{(N-1){A}_{N-1}}{4(N-3){(2 \pi   )}^{N-2}}}
{\frac{ \Gamma   ((N-1)/2) \zeta   (N-1)}
{{ \pi }^{(N+1)/2}}}{\frac{{\tilde{r}}_{+}^{N-3}}
{{\varepsilon }^{N-3}}} {\left({{\frac{{ \beta }_{  H}}
{{ \beta }^{}}}}\right)}^{  N-2}
\nonumber \\ 
&\times&
 \left[{{\frac{  (N-1)
(N-2)}{6}}  -{\frac{(N-1)(N-2)}{2}}\left({1-
{\frac{{r}_{-}^{N-2}}{{r}_{+}^{N-2}}}}\right)-{m}^{2}
{\tilde{r}}_{+}^{2}{ \sigma }^{2(b-c)}({r}_{+})}\right],
\end{eqnarray}
for $N\ne 3$, and
\begin{eqnarray}
{S}_{div2}&=&{\frac{1}{180}} {\left({{\frac{{ \beta }_{  H}}
{{ \beta }^{}}}}\right)}^{  3}  
\left[{2-{\frac{3}{1+{a}^{2}}}{\frac{{r}_{-}}{{r}_{+}}}}\right]\ln
\left({{\frac{{\Lambda }^{2}}{{\varepsilon }^{2}}}}\right)
\nonumber \\ 
&+&{\frac{1}{36}} {\frac{{ \beta }_{  H}}{{ \beta }^{}}}
  \left[{1-3\left({1-{\frac{{r}_{-}}{{r}_{+}}}}
\right)-3{m}^{2}{\tilde{r}}_{+}^{2}{\sigma }^{2(b-c)}({r}_{+})}
\right]\ln\left({{\frac{{\Lambda }^{2}}{{\varepsilon }^{2}}}}\right),
\label{last}
\end{eqnarray}
for $N=3$.

This result for $N=3$ can be compared with the results of other authors. 
The first line of Eq.~(\ref{last}) agrees with the evaluation of Ghosh
and Mitra \cite{7}, 
who also adopted the WKB method, when $a=1$ and in the limit $r_{-}=r_{+}$ 
after setting $\beta=\beta_{H}$.

The subleading contribution to the entropy has been evaluated also by 
Solodukhin \cite{8}. Their logarithmic terms for Schwarzschild and 
Reissner-Nordstr\"om black holes (Eqs.~(3) and (20) in 
the second paper in \cite{8}) 
coincide 
with the first term in (\ref{last}) with $a=0$. For a dilatonic black hole, 
although the 
expressions do not coincide exactly, the qualitative feature (its
positivity  in 
the extremal limit) is similar. The origin of the difference is
not clear up till now.

The second line in the expression (\ref{last}) has been found in this paper 
for the 
first time. This may not be a small contribution in comparison with the rest.
Although we have only given the first and second terms in the
expansion,  we would get the general feature of the series of
divergences. For a charged  dilatonic black hole (with $a\ne 0$) in the
extreme limit 
$r_{-}\rightarrow r_{+}$ (after setting $\beta=\beta_{H}$), the divergent 
quantum corrections as $\varepsilon \rightarrow 0$ would vanish except for a 
logarithmic term (for odd $N$), since higher divergent contribution
would be  written in positive powers of ${\tilde{r}}_{+}$.

To summarise, we have computed the leading and subleading divergence 
in the quantum correction to black hole entropy by using the WKB 
approximation. We have confirmed that the leading divergent piece of the 
entropy is proportional to the ``area'' of the horizon and has no
explicit dependence on other parameters describing the black hole. 
The subleading divergence obtained in this paper coincides partially with the 
known result. We have shown that there is a piece of the subleading 
divergence that comes from the next-to-leading term in the high-temperature 
expansion.


\end{document}